\documentclass[conference]{IEEEtran}
\IEEEoverridecommandlockouts
\usepackage{cite}
\usepackage{amsmath,amssymb,amsfonts}
\usepackage{algorithmic}
\usepackage{graphicx}
\usepackage{textcomp}
\usepackage{xcolor}
\usepackage{tabularray}
\usepackage[nolist]{acronym}
\usepackage[inline]{enumitem}
\usepackage[subrefformat=parens]{subcaption}
\usepackage{balance}
\usepackage{url}

\def\BibTeX{{\rm B\kern-.05em{\sc i\kern-.025em b}\kern-.08em
    T\kern-.1667em\lower.7ex\hbox{E}\kern-.125emX}}

\begin{document}

\begin{acronym}
    \acro{DCN}{Dynamic Circuit Network}
    \acro{DEP}{Digital Europe Programme}
    \acro{EQUO}{European QUantum ecOsystems}
    \acro{EU}{European Union}
    \acro{EuroQCI}{European Quantum Communication Infrastructure}
    \acro{IP}{Internet Protocol}
    \acro{KMM}{Key Management Module}
    \acro{KSID}{Key Stream ID}
    \acro{NBI}{North Bound Interface}
    \acro{ODF}{Optical Distribution Frame}
    \acro{OTDR}{Optical Time-Domain Reflectometer}
    \acro{PQC}{Post-Quantum Cryptography}
    \acro{QKD}{Quantum Key Distribution}
    \acro{QKDN}{Quantum Key Distribution Network}
    \acro{QNC}{Quantum Network Controller}
    \acro{QoS}{Quality of Service}
    \acro{SAE}{Secure Application Entity}
    \acro{SBI}{South Bound Interface}
    \acro{SDN}{Software Defined Network}
    \acro{TRL}{Technology Readiness Level}
    \acro{TTL}{Time To Live}
\end{acronym}

\title{European QUantum ecOsystems --- Preparing the Industry for the Quantum Security and Communications Revolution}

\author{
    \IEEEauthorblockN{
        Noel Farrugia\IEEEauthorrefmark{1},
        Daniel Bonanno\IEEEauthorrefmark{1},
        Nicholas Frendo\IEEEauthorrefmark{1},
        Andr\'e Xuereb\IEEEauthorrefmark{1},
        Evangelos Kosmatos\IEEEauthorrefmark{2}, \\
        Alexandros Stavdas\IEEEauthorrefmark{2},
        Marco Russo\IEEEauthorrefmark{3},
        Bartolomeo Montrucchio\IEEEauthorrefmark{3},
        Marco Menchetti\IEEEauthorrefmark{4},
        Davide Bacco\IEEEauthorrefmark{4}, \\
        Silvia Marigonda\IEEEauthorrefmark{5},
        Francesco Stocco\IEEEauthorrefmark{6},
        Guglielmo Morgari\IEEEauthorrefmark{6} and
        Antonio Manzalini\IEEEauthorrefmark{7}.
    }
    \IEEEauthorblockA{\IEEEauthorrefmark{1}Merqury Cybersecurity Limited, Malta}
    \IEEEauthorblockA{\IEEEauthorrefmark{2}OpenLightComm Europe, Czech Republic}
    \IEEEauthorblockA{\IEEEauthorrefmark{3}Politecnico di Torino, Italy}
    \IEEEauthorblockA{\IEEEauthorrefmark{4}QTI s.r.l., Italy}
    \IEEEauthorblockA{\IEEEauthorrefmark{5}Sparkle, Italy}
    \IEEEauthorblockA{\IEEEauthorrefmark{6}Telsy S.p.A, Italy}
    \IEEEauthorblockA{\IEEEauthorrefmark{7}TIM, Italy}
    
    \thanks{Corresponding author: N. Farrugia (email: noel@merqury.eu).}
    \thanks{© 2024 IEEE.  Personal use of this material is permitted.  Permission from IEEE must be obtained for all other uses, in any current or future media, including reprinting/republishing this material for advertising or promotional purposes, creating new collective works, for resale or redistribution to servers or lists, or reuse of any copyrighted component of this work in other works.}
}

\maketitle

\begin{abstract}
    There is mounting evidence that a second quantum revolution based on the
    technological capabilities to detect and manipulate single quantum particles
    (e.g., electrons, photons, ions, etc), a feat not achieved during the first
    quantum revolution, is progressing fast.
    It is expected that in less than 10 years, this second quantum revolution
    shall have a significant impact over numerous industries, including finance,
    medicine, energy, transportation, etc.
    Quantum computers threaten the status quo of cybersecurity, due to known
    quantum algorithms that can break asymmetric encryption, which is what gives
    us the ability to communicate securely using a public channel.
    Considering the world's dependence on digital communication through data
    exchange and processing, retaining the ability to communicate securely even
    once quantum computers come into play, cannot be stressed enough.
    Two solutions are available: Quantum Key Distribution (QKD) and Post-Quantum
    Cryptography (PQC); which, we emphasise, are not mutually exclusive.
    The EuroQCI initiative, of which EQUO is a part of, focuses on QKD and aims
    to build a network whereby EU countries can communicate securely through
    QKD\@.
    To this aim, the DEP (Digital Europe Programme) project aims to bring
    technological matureness to QKD by deploying a QKD test network and, through
    this exercise, understand what is lacking from an operator's point of view
    when the time to integrate QKD in their network comes.
\end{abstract}


\section{Introduction}

Cybersecurity is a term very commonly mentioned, and to good effect.
Without cybersecurity, the digital world we live in would cease to exist.
We can communicate, shop, and work remotely only thanks to our ability to
perform these tasks securely.
Remove the security, and everything breaks down, which is why it is of utmost
importance that, even if overlooked by many, investment in cybersecurity
research and development must continue.
Recently, a new threat unlike any before it has emerged: quantum computers.
Quantum computers have the capability, through Shor's algorithm~\cite{shor1994},
to break asymmetric cryptography in such a time frame to make it a very
perceivable threat.
Even though quantum computers powerful enough to break these encryption schemes
are not yet available, estimates of when these will be available range from five
to twenty years~\cite{mosca2022}.
Sensitive data that must remain secure for decades exists and includes
financial, governmental and health data amongst others.
Therefore, solving the cybersecurity threat posed by quantum computers is a
problem that must be tackled now due to the ``store now, decrypt later'' attack.
This is the sentiment shared by the \ac{EU} and the launch of the
\ac{EuroQCI}~\cite{euroqci} initiative.
The \ac{EuroQCI} initiative, which \ac{EQUO} is part of, is a push by the
\ac{EU} to develop and deploy a quantum communication network whereby keys
generated through \ac{QKD} can be shared between countries to secure the most
vital of information.
The role of \ac{EQUO} is to prepare the European \ac{QKD} industry to have high
\acp{TRL} (8--9) and high-performance components in preparation for the upcoming
large scale deployment of \ac{QKD} networks that needs to seamlessly coexist
with the already deployed classical network.
It is important to note that in \ac{EQUO}, both \ac{PQC} and \ac{QKD} are seen
as vital in the protecting the digital infrastructure with the most appropriate
technology being used in the most appropriate location~\cite{farrugia_2024}.
The technology developed in the \ac{EQUO} project will be demonstrated through
two deployments.
The first will be the deployment, management and control of a data centre link
in Israel.
The second is a \ac{QKD} network deployed in Turin's metropolitan area that will
test several ground-breaking features such as dynamic switching between a single
transmitter and multiple receivers.
More detail on the use cases can be found in
Section~\ref{sec:demos_and_use_cases}.
A comprehensive summary of the project is given in
Section~\ref{sec:project_summary}, followed by a detailed description of the
architecture used in \ac{EQUO} in Section~\ref{sec:functional_architecture}.
Section~\ref{sec:standardisation} summarises the standardisation efforts taken
on by \ac{EQUO} with Section~\ref{sec:conclusion} concluding this article.

\section{Project summary\label{sec:project_summary}}

\ac{EQUO} aims to develop and showcase the feasibility of discrete-variable
\ac{QKD} networks and their integration in current telecommunications
infrastructures as we pave the way for the Quantum Internet.
The architecture chosen by \ac{EQUO} follows the \ac{SDN} paradigm, where
network control is logically centralised.
As a matter of fact, the centralisation of control is the trend adopted for
current production telecommunication networks where multiple managers and
controllers are deployed for scalability and reliability reasons, which in turn
rely on distributed consensus protocols to operate in a logically centralised
manner.
Achieving the target \acp{TRL} (8--9) requires an architecture model that allows
the management and control of the \ac{QKDN}.
\ac{EQUO}'s architectural solutions are compliant (by design) to the most
relevant international recommendations (ITU-T, Y.38xx ``Quantum Key Distribution
Networks'' series) and European standards (ETSI, Group Specification documents
on \ac{QKD}), to maximise interoperability between modules in the same and
different \acp{QKDN}.

In summary, \ac{EQUO}'s vision, developments and experimental demonstrations are
expected to bring significant advances in the direction of the industrialisation
and deployment of EuroQCI, essential for the quantum protection of the European
digital infrastructure, and the development of an industrial European ecosystem,
including a thriving Small Medium Enterprise sector.

\section{Functional architecture\label{sec:functional_architecture}}

\subsection{Overview}

\begin{figure}[htbp]
    \centerline{\includegraphics[width=\linewidth,keepaspectratio]{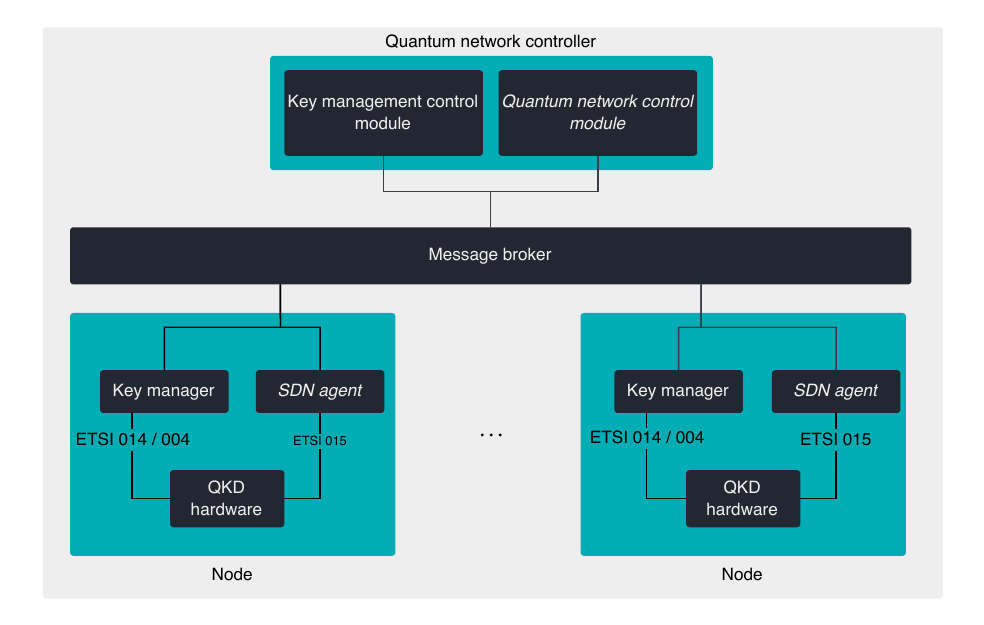}}
    \caption{
        High-level architecture used and developed in the \ac{EQUO} project.
        Blocks with italicised text are still in development.\label{fig:architecture}
    }
\end{figure}

The architecture used in the \ac{EQUO} project, shown in
Fig.~\ref{fig:architecture}, follows closely the one suggested by ITU-T
Y.3800~\cite{itut_y_3800}, whereby a logically centralised controller has full
view of the network topology and its current users.
An implementation choice we took as \ac{EQUO} worth highlighting is the use of a
message broker to act as the communication medium between the controller and the
nodes.
Multiple reasons were behind the decision to use a message broker, the most
important ones are summarised next.
The main reason is the separation of responsibilities.
Without a message broker the controller and nodes would have been responsible
for the successful delivery of messages, leading to a larger code base and the
development of a complex acknowledgement system.
With the use of a message broker, this responsibility is decoupled from the
service itself, be it controller or node and is shifted to the message broker
leading to a simpler code base for both the controller and nodes.
Such a separation of responsibility also allows for the network to be more
robust against sporadic losses of network connectivity.
The message broker employs a queueing system to buffer against intermittent
network issues.
For example, if the key management node is down due to a software update, the
controller can still send commands to the key management node.
The message broker queues this message, and it will be retrieved by the key
management service as soon as it is back online.
Such a mechanism is what allows the controller and node source code to be
simpler since the delivery burden is shifted on to the message broker.
Additionally, the use of a message broker gives us access to several packet
delivery methods, such as unicast, true multicast and broadcast.
These delivery methods are supported out of the box by the message broker and
are exploited by both the key manager and controller when operating and managing
the network.
Finally, messages can be sent using a node's service name rather than its
\ac{IP} address.
This makes development and management of the communication between multiple
services easier.
The introduction of a message broker may seem to increase the attack surface of
the quantum network.
However, if the necessary precautions are put in place, the security of the
network would be on the same level, if not easier to manage than if the
controller had to communicate directly with each node instance.
In \ac{EQUO}, a mutually authenticated and encrypted channel is used for
communication between a service and the message broker.
If needs be, the security of the messages broadcasted can be enhanced by using
\ac{PQC} along with the current classical algorithms and in later versions
encrypt the data using the keys generated through \ac{QKD}.

\subsection{Key management control module}

The controller responsible for servicing requests by the \ac{KMM} node and to
keep an up-to-date network topology of the network with real time information
and updates.
The latter is crucial in the ongoing work of supporting \ac{QoS} requests from
the \ac{KMM}.

\subsubsection{Topology discovery}

The \ac{KMM} is designed to communicate only with its peers; therefore, a single
\ac{KMM} instance does not have the full picture of the quantum network but only
information on who its neighbours are.
Therefore, on start-up, the \ac{KMM} transmits a \emph{hello-like} message to
the controller, including the ID of the \ac{KMM} sending the request for
identification, and the IDs of the neighbouring managers.
Once all \ac{KMM} services (one per location) are up, the network controller
has enough information to build the entire network topology.

\subsubsection{Path calculation}

In instances where an application requests a \ac{QKD} key for encryption but no
direct \ac{QKD} link between the two locations exists, the \ac{KMM} at the
source, asks the controller for a path connecting the two.
If a path exists, the shortest path in terms of hop count is calculated and
relayed to all the \ac{KMM}s on said path.
This gives enough information to the \ac{KMM} to proceed with key generation.
The shortest hop count path between two end nodes may not always be the best
path based on the network's current usage.
Work is progressing on improving the controller to take into consideration basic
\ac{QoS} constraints.
To be able to support this, a data model based off the ETSI GS QKD
015~\cite{etsi_qkd_015} standard will be agreed on with the \ac{KMM} where the
controller is updated on the \ac{KMM}'s key storage status.
Such information will be used by the controller during path calculation to avoid
the use of paths that are very low on key material, even if it means a longer
hop count.
Note that as the number of trusted nodes increases, so does the potential
risk of key exposure due to the increase in the number of trusted entities the
key is relying on to guarantee its security.
A quantum link status monitor service at the node is in development to notify
the controller and any other interested entities when a specific quantum link is
down or experiencing abnormal behaviour.
Thanks to the use of a message broker interested services will simply have to
subscribe to the notifications they are interested in.

\subsection{Node key manager\label{sec:node_key_management}}

Each \ac{KMM} module is responsible to support the following \ac{QKD}
functionalities:
\begin{enumerate*}
    \item
          Key storage
    \item
          Key protection
    \item
          Key provision to applications on request
    \item
          Key replacement on request
    \item
          Key destruction (based on decided key lifetime)
    \item
          Key Management
    \item
          Key reservation based on \ac{QoS} request
    \item
          Key relay to support a multi-hop network.
\end{enumerate*}
The \ac{KMM} serves requests from applications and returns keys under the
pre-specified \ac{QoS} performance along with overall management over the key
pool.
Key pool management ensures that all applications have enough key material and
removal of any expired keys.
Communication between the key management module and the network controller is
done through a message broker using a custom \ac{EQUO} specific interface.
The developed \ac{KMM} supports communication with the \ac{QKD} devices using
either the ETSI GS QKD 004~\cite{etsi_qkd_004} or ETSI GS QKD
014~\cite{etsi_qkd_014} standards.
The \ac{KMM} also includes a Key Allocation module which is responsible for the
allocation of the generated keys to the applications in line.
The Key Allocation module ensures that the allocation process is inline with:
\begin{enumerate*}
    \item the decided \ac{QoS} parameters, and
    \item the instructions received by the key management control module.
\end{enumerate*}
The \ac{KMM} is designed to work in a peer-to-peer fashion and synchronisation
with the other modules is done based off the KM link described in ITU-T
Y.3800~\cite{itut_y_3800}.
The key relay functionalities specified in ITU-T Y.3803~\cite{itut_y_3803} are
implemented to support key delivery that involves more than two \ac{KMM}
entities (multi-hop).
Finally, the \ac{KMM} includes a statistics manager for collecting and analysing
various statistics and a dashboard for the management of the \ac{KMM} including
a graphical presentation of the statistics.

\section{Demonstration and use cases\label{sec:demos_and_use_cases}}

\subsection{Use case 1 --- Tel-Aviv}

In this use-case, we demonstrate how to secure traffic using \ac{QKD} when it
transits through a Metropolitan Area Network in Tel-Aviv, to interconnect the
submarine cable landing station to a customer data centre.
The test bed involves the Burla Hub data centre where the marine cable coming
from Italy arrives to the data centre of Petah Tikva.
The two data centres are connected by a pair of dark fibre 15 kilometers in
length.
This fibre pair will be used for transmitting both the quantum and classical
channels (clock plus post processing communication).
The keys generated will be fed to two Telsy MusaX that will encrypt the traffic
generated by the Liqo application provided by Politecnico of Turin.
Such traffic will travel on a separate metro ring present in Tel-Aviv.

\subsubsection{Overview}

The metropolitan area of Tel-Aviv is chosen for the strategic role of this
region in interconnecting Europe and Asia.
In fact, many new data centres are currently being built here to provide
advanced services, with particular attention to security features.
Tel-Aviv is also a strategic point of landing for Mediterranean submarine cable
systems, e.g., BlueMed cable, connecting Italy to the Middle East and the new
high-strategic project of Blue Raman, that will connect Europe to Asia via
Middle East.
The central landing station of BlueMed/Blue Raman in Israel is the Burla data
center in Tel-Aviv, one of the two sites chosen for the \ac{EQUO} project.
The other site is Petah Tivka POP and Hub, located approximately 25 kilometers
away from Burla.
Petah Tivka is also the landing station for BlueMed and for the Mediterranean
Backbone cabling system (MedNautilus), a protected submarine ring-configured
backbone in the Mediterranean.

\subsubsection{Topology}

\begin{figure}[htbp]
    \centerline{\includegraphics[scale=0.7,keepaspectratio]{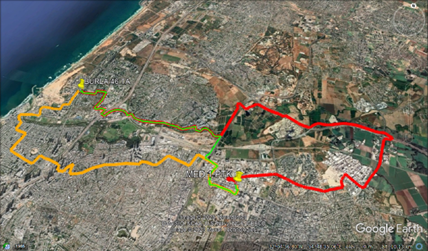}}
    \caption{EQUO fibre and metro ring paths.\label{fig:equo_fiber_and_metro_ring_paths}}
\end{figure}

Burla and Petah Tivka are connected by a metro ring, to ensure redundancy.
For the \ac{EQUO} project, Sparkle will provide a pair of single-mode dark
fibers between the two sites with the following characteristics: the first fibre
will be used for sharing the quantum key, and the second will be used for
synchronisation.
Finally, the metro ring (4 fibres in total: two couples of RX-TX) will be used
for encrypted traffic.
Routing of the dark fibre pair is different from the metro ring routing, but it
shares 80\% of the path with the northern path of metro ring.
The fibre's path is shown in Fig.~\ref{fig:equo_fiber_and_metro_ring_paths},
where the two parts of the metro ring are identified in red and orange while the
fibre dedicated to the \ac{EQUO} project is identified in green, highlighting
the overlapping region between the metro rings.
\begin{figure}[htbp]
    \centerline{\includegraphics[width=\linewidth,keepaspectratio]{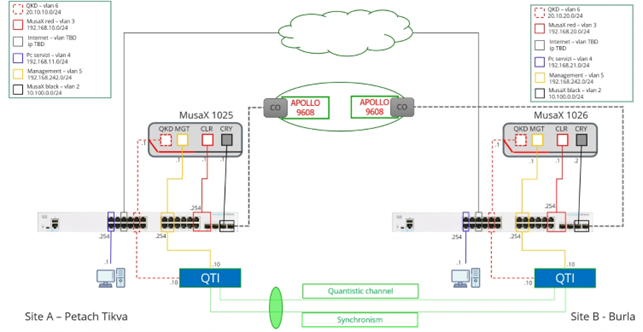}}
    \caption{Test bed architecture in detail.\label{fig:testbed_architecture_in_detail}}
\end{figure}
Fig.~\ref{fig:testbed_architecture_in_detail} provides a detailed overview of
the test bed architecture, including:
\begin{itemize}
    \item
          DWDM Apollo 9608 Ribbon directly connected to a Telsy MusaX.
          Apollo supports MPLS L2 on the internal side and L2 on cyphers side.
          Telsy MusaX integrates quantum keys provided by the \ac{QKD} system
          for encryption and uses a proprietary layer 3 VPN protocol called
          TelsyGuardX, designed to integrate \ac{QKD} and PQC primitives.
          MusaX periodically extracts the keys generated by the \ac{QKD} systems
          through the standard ETSI protocol (ETSI 014 standard) and uses the
          above keys to protect application-layer data.
    \item
          A number of Cisco switches for connecting the global system components
          (the keys interchange between QTI \ac{QKD} system and MusaX could be
          also direct and not through the switch)
    \item
          QTI \ac{QKD} and KME system for encryption (Quell-X) is a quantum random
          generator based on BB84 protocol.
          It supports ETSI 014 and SKIP standards and is equipped with a
          management and monitoring console based on TeamViewer to manage both
          the MusaX devices and the switches.
          Management interfaces are exposed over the Internet for a fully remote
          control.
    \item
          A computer running the Linux operating system connected to MusaX for
          traffic generation.
\end{itemize}

\subsubsection{OTDR Measurements}

An \ac{OTDR} is an instrument used to measure and create a visual representation
of a fibre optic cable route.
The measurement data can provide information on the condition and performance of
fibres, as well as any passive optical components along the cable path such as
connectors, splices, splitters, and multiplexers.
\ac{OTDR} measurement has been performed on the 14,919 meters of optical fibre
between Petah Tikva (A-end, port 55+56) and Burla (B-end, port 1-2) data
centres.
Measurement have been executed by Cellcom using Viavi \ac{OTDR} devices
considering the single pair of fibre (i.e., two fibres) and two measured
directions for each one (from A to B and from B to A).
Due to the presence of a ``patch panel'' between our cage in Petah Tikva station
and Cellcom, which is the only way to connect the Dark Fibre to our cage, there
are reflections on the \ac{OTDR} test related to this patch panel and its
connectors.
We did also several intervals of cleaning the fibres with scope and replacing
all patch-cords with new ones.
\ac{EQUO} project requires an attenuation \textless{} 20dB and by \ac{OTDR}
instrument we have measured an attenuation value of about 0.3 dB/km between the
two data centres.

\subsection{Use case 2 --- Turin metropolitan area}

\begin{figure}[htbp]
    \centerline{\includegraphics[width=\linewidth,keepaspectratio]{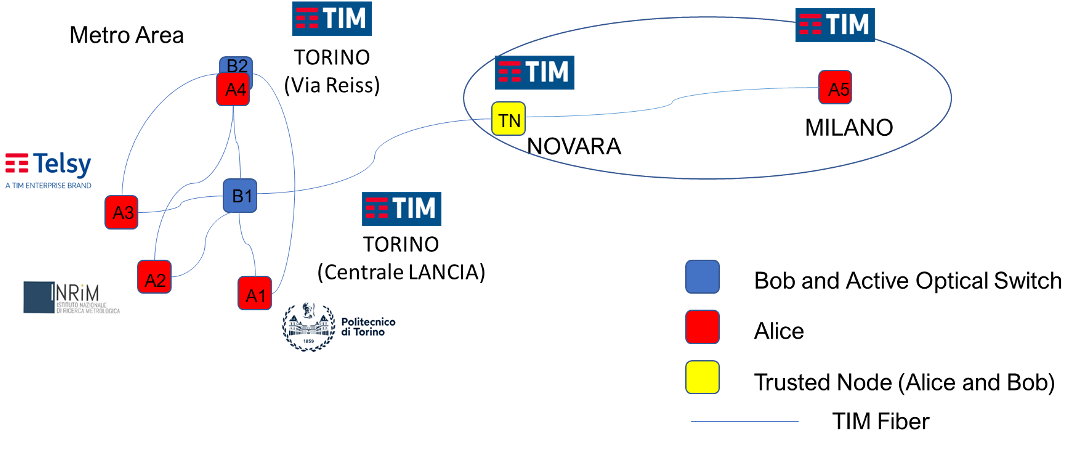}}
    \caption{Turin use case.\label{fig:turin_use_case}}
\end{figure}

The Turin use case aims to demonstrate advanced functionalities of the \ac{QKD}
network, beyond point-to-point links, in the metropolitan network of Torino,
Italy.
\ac{QKD} nodes will be installed in the locations of TIM, Telsy, INRIM and the
Labs of the Politecnico of Turin as shown in Fig.~\ref{fig:turin_use_case}.
The network topology design is mostly dictated by the geography and the tests
and demonstrations that have to be carried out as part of the project.
This network will have \ac{SDN} capabilities, through standard protocols,
interfaces, and optical switching of the quantum channel, which will allow
advanced quantum cyber-security services (e.g., time-sharing the quantum link
between one Bob, multiple Alices) seamlessly integrated in the metropolitan
network.
In one instance of the use-case, the traffic generated by a distributed
computing platform running at the Politecnico of Turin and transported to TIM
locations will be protected by \ac{QKD} services, thus showing a seamless
integration of \ac{QKD} with the fog-edge computing paradigm.
In another instance of the use-case, data traffic generated by measurements at
INRIM will be protected by \ac{QKD} across links reaching TIM and Telsy
locations.
Other instances are under definition to show advanced \ac{QKD} services beyond
point-to-point.
The experiments will start in the second half of 2024 with the results to be
shared in another publication.

\subsection{Application layer}

According to recommendation ITU-T Y.3802, the application layer supports the
Cryptographic application function; it consumes the shared key-pairs provided by
a \ac{QKDN} and performs secure communication between remote parties.
Telsy's ciphers interact with the respective \ac{KMM} using the ETSI GS QKD
014-based protocol.
The communication between different ciphers, including Key ID notification,
follows a proprietary protocol.
Data traffic is generated by a distributed application named Liqo that will run
on the cryptographic layer provided by Telsy's encryption modules.
Liqo is an open-source project that enables dynamic and seamless Kubernetes
multi-cluster topologies, supporting heterogeneous on-premises, cloud and edge
infrastructures.
Two Telsy encryption modules receiving the same key from the \ac{KMM} are
capable of establishing a secure channel, which carries the Liqo application
data.
This channel implements a transport protocol, whose security is based on the
aforementioned shared key, integrated with public key cryptography mechanisms.
The resulting combination ensures the integrity, authenticity, and
confidentiality of the data, in addition to the more subtle properties required
by modern cryptographic paradigms (e.g., perfect forward secrecy).
The refresh time of the shared key can be modified according to the Liqo
application needs; the higher the data rate, the more frequent the key refresh.

\balance{}

\section{Bridging the standardisation gaps\label{sec:standardisation}}

As mentioned earlier, one of the aims of the \ac{EQUO} project is the
identification, and solution where possible, of gaps that are hindering the
uptake of \ac{QKD} within the telecommunication industry.
The lack of standards is one such hindrance we aim to overcome.
The lack of a specific key retrieval protocol is one such gap.
Both the ETSI GS QKD 014 and 004 standards have been designed as an interface
between the application, referred to as the \ac{SAE}, and the key management
system.
And thus, both protocols are based on the application requesting keys from the
devices.
While this design choice is suitable between an application and a key manager,
it is not suitable to interface between the key manager and the \ac{QKD} device
because the key manager will have to poll the device periodically for key
material.
To this extent we see the need of a push-based protocol where the \ac{QKD}
device pushes the key to the key management system once it is ready.
This increases the link's efficiency interconnecting the key manager with the
\ac{QKD} device.
After analysing the current status quo and speaking with \ac{QKD} manufacturers,
including QTI who is part of the \ac{EQUO} consortium, it became apparent that
either a new standard or modification to current ones are required.
To avoid work duplication, it has been decided that modifications to the current
ETSI GS QKD 004 standard is the best and fastest way to get this new push based
protocol standardised.
The availability of a push protocol means that the key manager does not have to
periodically poll the \ac{QKD} hardware for keys, but rather listen to
notifications by the \ac{QKD} device.
This ensures that the key manager has the keys in storage as soon as they are
made available and reduces the unnecessary and inefficient polling load from the
\ac{QKD} hardware.
Finding an optimal polling frequency is not a trivial problem to solve
considering that the quantum link is very dynamic and unpredictable.

\section{Conclusion\label{sec:conclusion}}

Throughout this publication the importance of having a method for quantum secure
communications, based on \ac{QKD}, has been highlighted.
This is also evidenced by the European Union's push geared towards cybersecurity
research, of which the \ac{EuroQCI} initiative has been highlighted here.
\ac{EuroQCI} is a framework whereby the EU member states research, learn and
develop technologies related to \ac{QKD}.
\ac{EQUO}, is a \ac{DEP} project, that falls under the \ac{EuroQCI} industrial
umbrella targeted towards the preparation of \ac{QKD} integration by the telco
industry, as evidenced by the broad range of expertise that make up \ac{EQUO}'s
consortium.
To demonstrate the capability of \ac{QKD}'s deployment in a network, two use
cases have been presented.
One being the protection of data centre traffic in a transmission ring in Tel
Aviv (Israel) and the other being a metropolitan \ac{QKD} network in Turin aimed
to demonstrate the dynamic nature of a \ac{QKD} network.
Standardisation is key for the uptake of \ac{QKD} technology by integrators and
network operators, with the efforts being made by the \ac{EQUO} consortium in
the area highlighted.

\section*{Acknowledgement}
This paper has been developed in the project EQUO (European QUantum ecOsystems)
which is funded by the European Commission in the Digital Europe Programme under
the grant agreement No 101091561.

\bibliography{references}

\end{document}